\documentclass[final,3p,times]{elsarticle}

\usepackage{float}
\usepackage{array}
\usepackage{tabularx}
\usepackage{subfigure}
\usepackage{colortbl} 
\usepackage{setspace}
\usepackage{url}
\usepackage{amsmath, bm}
\usepackage{stfloats}
\usepackage{graphicx}
\usepackage{hyperref}
\usepackage{gensymb}
\usepackage{booktabs}
\usepackage{multicol}
\usepackage{multirow}
\usepackage{makecell}
\usepackage{amssymb}
\usepackage[table]{xcolor}
\definecolor{gray}{rgb}{.949,.949,.949}
\newcolumntype{Y}{>{\centering\arraybackslash}X}
\hypersetup{hidelinks,colorlinks=false,pdfstartview=Fit,breaklinks=true}

\usepackage{algcompatible}
\usepackage{algorithm}
\usepackage{color}
\usepackage{url}
\usepackage{subfigure}

\usepackage{amssymb}

\begin{document}

\begin{frontmatter}

\title{Coordinated Power Smoothing Control for Wind Storage Integrated System with Physics-informed Deep Reinforcement Learning}

\author[inst1,inst2]{Shuyi Wang}
\author[inst3]{Huan Zhao}
\author[inst1]{Yuji Cao}
\author[inst1,inst2]{Zibin Pan}
\author[inst4]{Guolong Liu}
\author[inst5]{Gaoqi Liang}
\author[inst1,inst2]{Junhua Zhao}
\affiliation[inst1]{organization={The Chinese University of Hong Kong, Shenzhen},
            postcode={518172}, 
            country={China}}
\affiliation[inst2]{organization={Shenzhen Institute of Artificial Intelligence and Robotics for Society (AIRS)},
            postcode={518000}, 
            country={China}}
            
\affiliation[inst3]{organization={The Hong Kong Polytechnic University},
            postcode={710048}, 
            country={China}}
            
\affiliation[inst4]{organization={Nanyang Technological University},
            postcode={639798}, 
            country={Singapore}}

\affiliation[inst5]{organization={Harbin Institute of Technology, Shenzhen},
            postcode={518055}, 
            country={China}}

\begin{abstract}
The Wind Storage Integrated System with Power Smoothing Control (PSC) has emerged as a promising solution to ensure both efficient and reliable wind energy generation. However, existing PSC strategies overlook the intricate interplay and distinct control frequencies between batteries and wind turbines, and lack consideration of wake effect and battery degradation cost. In this paper, a novel coordinated control framework with hierarchical levels is devised to address these challenges effectively, which integrates the wake model and battery degradation model. In addition, after reformulating the problem as a Markov decision process, the multi-agent reinforcement learning method is introduced to overcome the bi-level characteristic of the problem. Moreover, a Physics-informed Neural Network-assisted Multi-agent Deep Deterministic Policy Gradient (PAMA-DDPG) algorithm is proposed to incorporate the power fluctuation differential equation and expedite the learning process. The effectiveness of the proposed methodology is evaluated through simulations conducted in four distinct scenarios using WindFarmSimulator (WFSim). The results demonstrate that the proposed algorithm facilitates approximately an 11\% increase in total profit and a 19\% decrease in power fluctuation compared to the traditional methods, thereby addressing the dual objectives of economic efficiency and grid-connected energy reliability.
\end{abstract}


\end{frontmatter}


\section{Introduction}
\label{sec:sample1}
As one of the most popular renewable energy resources, wind power holds substantial potential for meeting future global energy demands while mitigating climate change and environmental pollution. However, the intermittent nature of wind power introduces inherent variability and uncertainty when integrated into power systems. As the wind power penetration level increases, the secure and reliable operation of power systems becomes a significant challenge \cite{yin2023multi}. In practice, the grid usually requires the active power fluctuation from wind farms to be confined to a specific value within a one-minute time window \cite{shivashankar2016mitigating}. 
Therefore, Wind Power smoothing control (PSC) has emerged as a potential solution. 
Previous research has established two major categories of Power Smoothing Control for wind farms, including regulation control of wind turbines and indirect power control by Battery Energy Storage System (BESS). 
The former approach typically involves pitch angle control \cite{tang2018active}, rotor inertia control \cite{kim2016power}, and Direct Current (DC)-link voltage control \cite{uehara2011coordinated}, which require a different operation from maximum power point tracking, causing inefficiency and potential damages \cite{barra2021review}. 

On the contrary, with a stronger capability of power smoothing, the BESS-based PSC coordinates the active power from BESS and wind turbine \cite{lu2024advances}, providing rapid response to power fluctuation with high operability and little power loss. Recognizing the benefits of such Wind Storage Integrated Systems (WSIS) \cite{zhang2020accommodate}, incentive policies have been introduced to mandate the installation of BESSs from 10\% to 30\% of wind farms’ installed capacity. WSIS facilitates wind power storage, allocating, and smoothing, enhancing delivery stability and energy management flexibility for both the grid and wind farm.

For BESS-based power smoothing, traditional model-based methods utilize the physical properties of WSIS to build the models and optimize the power generation. In terms of model completeness, researchers have considered the wind turbine wake model \cite{wang2023coordinated_tse}, battery degradation model \cite{huang2019hierarchical}, wind speed and power forecasting model \cite{wang2023coordinated}, and other relevant physics information for PSC problems in WSIS. Moreover, Xiong et al. \cite{xiong2021optimal} combine sophisticated wake and battery energy models to optimize the allocation of power flow in BESS. However, challenged by complexity, none of the studies include both the wake effect and battery degradation in the model formulation of the WSIS PSC problem. 
Such model-based approaches overly rely on the accuracy and temporal scale of the environmental models, necessitating a trade-off between control deviation and computational costs, which pose difficulties for real-time implementation.

To address the aforementioned challenges, several model-free methods are applied in WSIS, such as random search \cite{ahmad2016model}, adaptive control \cite{dong2020model}, and Reinforcement Learning (RL). As an emerging data-driven approach, RL allows the agent to learn optimal policy by performing actions and receiving feedback rewards from the environment, making it well-suited for complex, dynamic, and uncertain conditions \cite{sutton2018reinforcement}. Deep RL (DRL), which incorporates RL with deep neural networks \cite{cao2024survey}, has been successfully demonstrated in wind farm and BESS, respectively. For instance, Deep Deterministic Policy Gradient (DDPG) algorithm has been applied to achieve optimal wind power generation \cite{zhao2020cooperative} and power smoothing \cite{zhu2022optimal} in wind farms. Yang et al. \cite{yang2020deep} implemented the Rainbow algorithm to control the charge/discharge schedule of the BESS to increase their revenues under uncertainties of wind generation and electricity price. 
Furthermore, with a unitary control frequency, Wang et al. \cite{wang2023coordinated_tse} proposed a hybrid power smoothing strategy for WSIS. This approach consists of a model-based power control and an RL-based power optimization.
However, there is currently a scarcity of research that fully addresses the WSIS PSC problem using model-free methods, especially satisfying both power maximization and power smoothing. Moreover, previous coordinated PSC studies have overlooked the distinct response frequencies and internal control sequences of wind turbines and BESS, which is a pivotal factor to take into account.

To navigate the intricate WSIS environment and enhance the learning process efficiency, multi-agent RL (MARL) and Physics-informed Neural Network (PINN) are introduced.
Based on game theory, MARL deals with self-learning and decision-making in a dynamic environment where multiple agents interact, cooperate, or compete with each other based on local observations \cite{bucsoniu2010multi}. As a result, MARL can be introduced to handle asynchronous decision-making and multi-objectives. Furthermore, multi-level MARL explores settings where agents are organized into hierarchies or teams \cite{feng2022multi}, each with different responsibilities and coordination requirements, which allows for much more flexible task allocation. MARL has been applied to manage multiple wind turbines \cite{wang2023coordinated_tse} and batteries \cite{dada2021application} for decision optimization in smart grid. 
On the other hand, PINN combines traditional physics-based modeling with machine learning methods \cite{karniadakis2021physics}, which makes it possible to design specialized network architectures that automatically satisfy some of the physical invariants for better accuracy, faster training, and improved generalization. Specifically, several paradigms of PINN, e.g., Physics-informed (PI) loss function, PI initialization, PI design of architecture, and hybrid physics-deep learning models, are summarized in \cite{huang2022applications}. PINN has found diverse extensions and applications, including dynamic analysis \cite{stiasny2021learning} and transfer learning \cite{chakraborty2021transfer}. When cooperating physics priors into the RL, it highlights the different components of the typical RL paradigm, such as state/action space, reward function, and agent networks \cite{banerjee2023survey}. Gao et al. \cite{gao2022transient} proposes a transient voltage control approach, by using the transient constraint on the value function to expedite convergence. However, for dynamic complex and uncertain environments, further exploration is imperative regarding the selection of both the physics prior and the method of RL implementation.

To solve the PSC problem in WSIS, this paper proposes a Physics-informed Deep Reinforcement Learning (PIDRL)-based coordinated control framework. The contributions of this paper are summarized as follows:
\begin{itemize}
    \item In light of the distinct control frequencies and sequence of wind turbines and BESS, a bi-level coordinated control framework is derived to address the WSIS PSC problem, which also incorporates wake effect and battery degradation models to achieve a comprehensive and authentic environment.
    \item The proposed coordinated PSC framework is reformulated to a bi-level Markov decision process, and the reward function of MARL is redesigned to derive the real-time optimal strategy with less uncertainty. Specifically, the power generation of the wind farm, power fluctuation to the grid, energy loss within BESS, the degradation cost of BESS, and the penalty of the unsafe actions are selectively constructed as the reward functions of the two agents, to maximize total profitability and ensure power smoothness.
    \item A physics-informed neural network in a MARL-based control algorithm is integrated into actor-network training to accelerate the learning process. A customized partial differential equation about power fluctuation is conceived as a weighted loss term within the network architecture of DDPG. The proposed PINN-assisted Multi-agent DDPG (PAMA-DDPG) method has showcased superior performance for the WSIS PSC problem.
\end{itemize}

The remaining sections of this paper are organized as follows. In Section \ref{Section II}, we briefly describe the WSIS models. In Section \ref{Section III}, the PIDRL-based coordinated WSIS PSC framework is introduced, and the PAMA-DDPG algorithm is proposed. Section \ref{Section IV} verifies the effectiveness of the proposed method under several scenarios in WindFarmSimulator (WFSim). Section \ref{Section V} gives the conclusion and future works of this paper.

\section{Wind Storage Integrated System}
\label{Section II}
\subsection{Overall Structure}
\begin{figure}[htpb]
    \centering 
    \includegraphics[width = 0.6\linewidth]{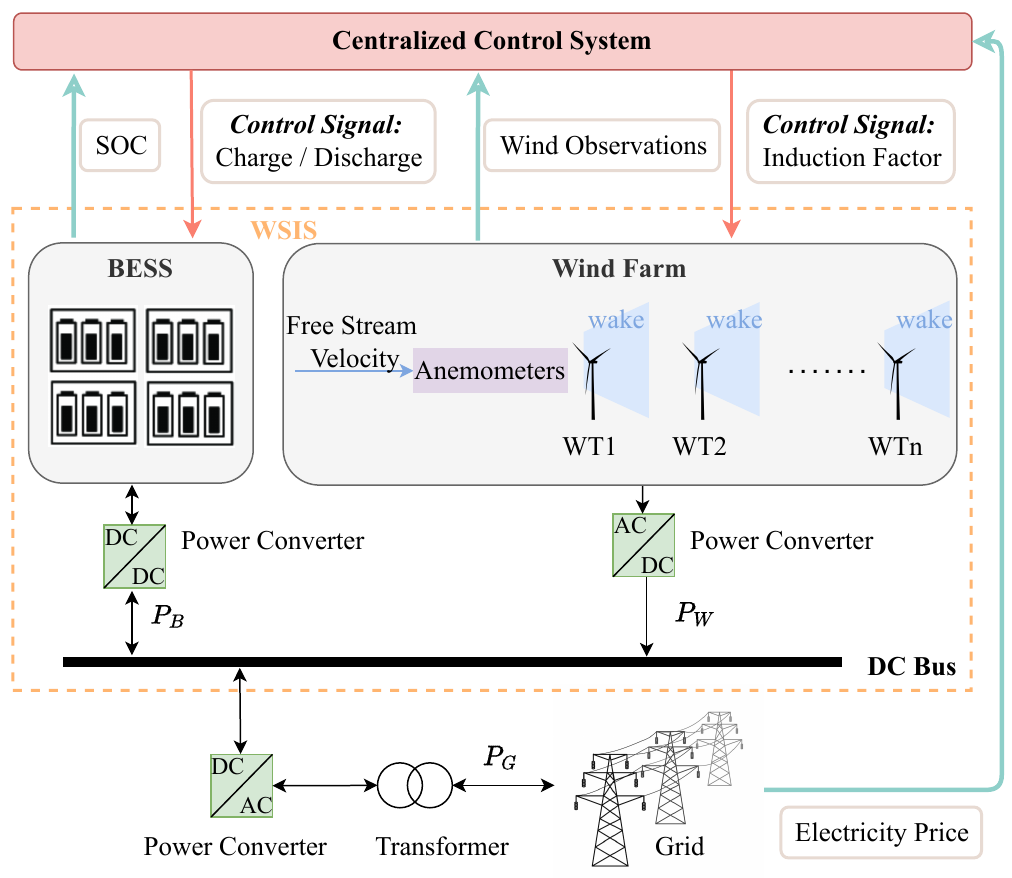}
    \caption{The structure of the wind storage integrated system.} 
    \label{f1}
\end{figure}

The WSIS primarily consists of a centralized control system, the BESS, wind turbines, and transmission lines that connect the system to the main grid. Power converters and transformers serve as devices for power conversion. The overall structure of the WSIS is shown in Fig. \ref{f1}. 

Within the wind farms, anemometers are employed to measure the free-stream velocity, including both wind speed and direction. Based on the induction factor received from the centralized control system, the turbines capture the kinetic energy from the wind and convert it into electrical energy, where the wake effect impacts the downstream wind turbines by reducing wind speed and generating additional turbulence.
The BESS is connected to the DC bus, allowing for bidirectional power flow. It effectively stores or dispatches power according to the control signal, ensuring a reliable power supply and maximizing revenue. 
The centralized control system serves to optimize the operation of the wind turbines and manages the power flow of BESS, thereby ensuring grid stability through continuously monitoring the observations, including the State of Charge (SOC), wind velocity, and electricity price. During periods of excess wind generation, surplus power can be fed into the BESS. Moreover, BESS may also be required to charge from the grid when the electricity price and SOC are relatively low. Conversely, when there is a shortfall in wind power generation, stored energy from the BESS can be utilized to smooth the fluctuations.

\subsection{Wind Farm Model}
\subsubsection{Wind Turbine Model}
The power generated by a single wind turbine can be mathematically formulated using the following equations:
\begin{equation}
\label{wind_0}
P_{turbine}=\frac{1}{2}\ \rho A_rC_pU_{inf}^3
\end{equation}
\begin{equation}
\label{cut-wind}
U_{min} \le U_{inf} \le U_{max}
\end{equation}
where $P_{turbine}$ is the power output of the wind turbine, $\rho$ is the air density, $A_r$ is the swept area of the rotor, and $U_{inf}$ is the free stream speed. The power coefficient, $C_p$, is the ratio of electric power extracted from the wind flow and is explicitly concerning detailed control policy. Furthermore, free stream speed $U_{inf}$ should be in a threshold between the cut-in speed $U_{min}$ and cut-out speed $U_{max}$ for regular operation of the turbine. 

The Actuator Disk Model (ADM) \cite{mikkelsen2003actuator} is used to simulate a wind turbine's aerodynamics, where the wind turbine's rotor is regarded as an actuator disk. Applying ADM
to a one-dimensional stream shows that:
\begin{equation}
\label{wind_1}
    C_p \left(\alpha\right)
    = 4\alpha\left(cos \left(\varphi_y\right)- \alpha\right)^2
\end{equation}
\begin{equation}
    \alpha =1-\frac{U_r}{U_{inf}}
\end{equation}
where $\varphi_y$ is the yaw angle and $U_r$ is the wind speed at the rotor. $\alpha$ $(0\le\alpha\le0.5)$ is the axial induction factor, which is defined as the ratio of the speed deficit behind the rotor to the undisturbed wind speed ahead of the rotor. In axial induction-based control, the axial induction factor of the turbines is tuned by physically controlling the blades’ tip speed ratio and pitch angle to recover the flow \cite{annoni2016analysis}.

\subsubsection{Wake Model}
The wake effect introduces challenges in calculating power output, as the turbulence and reduced wind speeds in the wake impact the downstream wind turbines. The Jensen Park model \cite{katic1986simple} takes into account the wake expansion and the effect of the rotor on the incoming wind flow. The wind speed $U$ at location $(x,y)$ is defined as:
\begin{equation}
\label{wind_3}
    U\left(x,y,\alpha\right)=U_{inf}\left(1-u_{deficit\ }\right)
\end{equation}
where $u_{deficit}$ is the wind deficit percentage compared to the free stream speed $U_{inf}$, represented by: 
\begin{equation}
\label{wind_4}
u_{deficit\ } = \left\{
\begin{aligned}
&\frac{2\alpha}{\left(1+\frac{2kx}{D}\right)^2},       &      & \mbox{if } y\le\ \frac{D+2kx}{2}\\
&0,     &      & \mbox{otherwise } \\
\end{aligned} \right.
\end{equation}
where $D$ is the turbine blade's diameter, and $k$ is the wake expansion parameter. Due to its comprehensibility and broad applicability, the Jensen Park model is commonly used in mainstream wind farm simulators.

\subsection{Battery Energy Storage System}
\subsubsection{Battery Energy Model}
For the purpose of simulation, the dynamics of the BESS considering energy loss \cite{li2019constrained} is modeled as:
\begin{equation}
\label{BESS_1}
   E_B^{t+1} =E_B^t+[{max(P_{B}^t,0)} \cdot \eta_{ch}-\frac{{max(-P_{B}^t,0)}}{\eta_{dis}}] \cdot \Delta t 
\end{equation}
where $\eta_{ch}$ and $\eta_{dis}$ represent the energy conversion efficiency during the charging and discharging process, respectively. $\Delta t$ is the duration time of each interval and $E_B^t$ is the battery energy at time $t$. $P_{B}^t$ represent the power flow to BESS, wherein the $P_{B}^t>0$ and $P_{B}^t<0$ denote the processes of charging and discharging. Note that  the maximum charging and discharging rates at which the battery can transfer electrical energy \cite{zhao2022mobile} are:
\begin{equation}
\label{BESS_2}
    -P_{max}^{dis} \leq P_{B}^t \leq P_{max}^{ch}
\end{equation}
where $P_{max}^{ch}$ and $P_{max}^{dis}$ are the power limits of charging and discharging, respectively. Besides, to ensure the safe and reliable operation of the BESS, the constraint for battery energy is as follows:
\begin{equation}
\label{BESS_3}
    {E_{min}\le E}_B^t\le E_{max}
\end{equation}
where $E_{min}$ and $E_{max}$ denotes the minimum and maximum energy levels, respectively.

\subsubsection{Battery Degradation Model}
Battery degradation affects the energy that can be stored and discharged, ultimately impacting the system's reliability and economics. A linear programming approach in \cite{bordin2017linear} is used to model and estimate the equivalent battery degradation costs. Firstly, the lifetime throughput $L_{T,n}$ for every depth of discharge $n$ is calculated as:
\begin{equation}
    L_{T,n}=E_{max}\cdot g_n \cdot f_n
\end{equation}
where $f_n$ is the number of battery cycles to failure, $g_n$ is the Depth of Discharge (DOD) of the battery. Then, the resulting lifetime throughput $L_T$ is obtained by averaging the values of $L_{T,n}$ in the allowable operating range of battery DOD:
\begin{equation}
    L_T=\frac{1}{n}\sum_{1}^{n}L_{T,n}
\end{equation}
The equivalent battery degradation cost per kWh can be defined as:
\begin{equation}
    K_{deg} =\frac{C_{R}}{L_T\cdot RE}
\end{equation}
where $C_{R}$ is the replacement cost of the battery, and $RE$ is the square root of the roundtrip efficiency of the battery. The unit of $K_{deg}$ is $\left(\$/{kwh}\right)$. This modeling approach refers to battery lifetime throughputs under different DOD and SOC and performs averaging to estimate the battery degradation cost in a linearized way. The cost will be incurred every time the battery is discharged:
\begin{equation}
\label{BESS_4}
    C_{deg}^t= K_{deg} \cdot {max(-P_{B}^t,0)}\cdot \Delta t
\end{equation}

\section{DRL-based Coordinated Control Framework}
\label{Section III}
\subsection{Bi-level Coordinated Power Smoothing Control}
The objective of the proposed framework is to maximize the total power profit of the WSIS while ensuring the fluctuation of the power output within an acceptable level. The key idea of this framework is based on two assumptions. 
First, the centralized control system has the ability to control all the turbines simultaneously.
Second, the control frequencies for wind turbines and the BESS are set to be $1/f$ and $1$ $\rm {min}^{-1}$, respectively, where $f$ is the real-time control time scale for wind turbines  \cite{cortina2017investigation}. According to the highest control frequency, the time granularity $\Delta t$ is discretized as one minute. 

The model of the PSC problem for WSIS is formulated as an objective function (\ref{obj}) that maximizes the total profit of the power generation and the negative degradation cost of the BESS in a time interval of $T$:
\begin{equation}
\label{obj}
    \begin{split}
        \max\limits_{\vec{\alpha}^t,P_{B}^t } &{\sum_{t=0}^{T}{(Pr^t\cdot P_{G}^t \cdot \Delta t}-C^t_{deg})}
    \end{split}
\end{equation}
s.t. 
\begin{equation}
    \vec{\alpha}^t=\left(\alpha_1^t,\alpha_2^t,\ldots,\alpha_n^t\right)
\end{equation}
\begin{equation}
    P_W^t = \sum\limits_{i \in \mathcal{N}_\Psi} P_{turbine,i}^t
\end{equation}
\begin{equation}
\label{flow_eq}
    P_W^t-P_B^t=P_G^t
\end{equation}
\begin{equation}
\label{power_fluc}
    P_{FG}^t=\left|P_G^{t-1}-P_G^t\right|
\end{equation}
\begin{equation}
\label{power_fluc_cons}
     P_{FG}^t \le P_{FG, max}
\end{equation}
$$  \mbox{Eqs. (\ref{wind_0}) - (\ref{BESS_3}), (\ref{BESS_4})} $$
where $Pr^t$, $P_G^t$, $\vec{\alpha^t}$, and $P_W^t$  denote the electricity price, the power connected to the grid, the vector of the turbines' induction factors, and the power generated by wind farm. $\mathcal{N}_\Psi$ is the set of wind turbines with size $n$ and $P_{turbine,i}^t$ is the wind power of turbine $i$ in time $t$ under the wake effect. $P_{FG}^t$ is the power fluctuation to grid and $P_{FG,max}^t$ is its threshold.
Eq. (\ref{flow_eq}) is the dynamic equation of the power flow in WSIS, where $P_G^t>0$ and $P_G^t<0$ indicate power injection into and extraction from the grid, respectively.
Eq. (\ref{power_fluc}) defines the grid-connected power fluctuation $P_{FG}^t$ as the changes of the $P_G^t$ over $\Delta t$.
Eqs. (\ref{wind_0}) - (\ref{wind_1}) denotes the model of a single wind turbine. Eqs. (\ref{wind_3}) - (\ref{wind_4}) consider the wake model. Eqs. (\ref{BESS_1}) - (\ref{BESS_3}) model the energy dynamics of the BESS and Eq. (\ref{BESS_4}) consider the battery degradation cost through a coefficient $K_{deg}$.

To effectively tackle the separate control frequencies between wind turbines and BESS within the context of the WSIS problem, a bi-level formulation is proposed. For clear separation of concerns and reduction of complexity, each level is set to be in charge of the turbine-related and BESS-related objectives and constraints, respectively. By employing suitable optimization algorithms and techniques, each level can be solved successively, thereby simplifying the computational burden. 
Given that BESS operates at a much higher frequency and relies on the decisions made by the wind turbines, the upper level is designed to focus on the wind turbines, while the lower level is dedicated to the BESS. This bi-level formulation helps optimize the overall system operation and achieve a balance between responsiveness and efficiency. It ensures that the wind power generation is maximized at the upper level, while the lower level effectively manages the power smoothing and reduces the degradation cost.

\subsubsection{Upper-level Wind Farm Control}
According to the separation of the bi-level formulation, the sub-goal of the upper level is to maximize the total profit of wind power generation in a control loop $T$ response to changes in wind velocity and electricity price. Choosing all turbines' axial induction factors $\vec{\alpha}$ as the control variable, the control objective can be written as: 
\begin{equation}
    \max\limits_{\vec{\alpha}^t} {\sum_{t=0}^{T}} {\left(Pr^t\cdot {\sum_{i\in \mathcal{N}_\Psi} {P_{turbine,i}^t}}\right)}
\end{equation}
s.t.
\begin{equation}
\label{wind_control_freq}
    \vec{\alpha}^t = \vec{\alpha}^{t-1} \quad \mbox{if not $t=0,f,2f,...$}
\end{equation}
$$  \mbox{Eqs. (\ref{wind_0}) - (\ref{wind_4})} $$
The control frequency is guaranteed in Eq. (\ref{wind_control_freq}). The detailed models for wind turbines are illustrated in Eqs. (\ref{wind_0}) - (\ref{wind_4}).

\subsubsection{Lower-level Battery Energy Storage System Control}
The lower-level system is related to control objects in BESS, which are responsible for minimizing the battery degradation cost, electricity purchasing cost, and power fluctuations by controlling the power flow among the wind farm, BESS, and the grid. For simplicity and flexibility, the power fluctuation constraint Eq. (\ref{power_fluc_cons}) is transferred to the objectives as a penalty term that prescribes a high cost for violation of the constraint. Therefore, the problem is represented as:
\begin{equation}
\label{obj_l}
    \min\limits_{P_{B}^t} {\sum_{t=0}^{T}} \left[ {C_{deg}^t + Pr^t\cdot(P_B^t-P_W^t)+\beta\cdot P^t_{VG}} \right]
\end{equation}
s.t.
\begin{equation}
\label{penalty_vio}
    P^t_{VG}=\left\{
    \begin{aligned}
&P^t_{FG},     &   &    \mbox{if } P_{D}^t \leq 0\\ 
&P_{FG, max}+\nu\cdot P_{D}^t ,       &      & \mbox{otherwise}\\
    \end{aligned} \right.
\end{equation} 
\begin{equation}
    P_{D}^t = P^t_{FG}-P_{FG, max}
\end{equation}
$$  \mbox{Eqs. (\ref{BESS_1}) - (\ref{BESS_3}), (\ref{BESS_4}), (\ref{power_fluc})} $$
where $P^t_{VG}$ is defined as the power violation penalty and $P_D^t$  signifies the surplus beyond the threshold value. $\beta$ and $\nu$ are coefficients related to the penalty terms. In order to account for the discrepancy in units between violation and costs, it is necessary to scale and adjust their respective impacts on the control task by the penalty coefficient $\beta$, which effectively modifies their relative influence. To more effectively penalize power fluctuations beyond the specified threshold, denoted as $P^t_{D}$, the violation penalty $P^t_{VG}$ is formulated as a piecewise function with a relatively large coefficient $\nu$. The detailed models for BESS are illustrated in Eqs. (\ref{BESS_1}) - (\ref{BESS_3}) and (\ref{BESS_4}).

\subsection{DRL-based Coordinated Control Framework}

\begin{figure}[t!]
    \centering 
    \includegraphics[width = 0.6\linewidth]{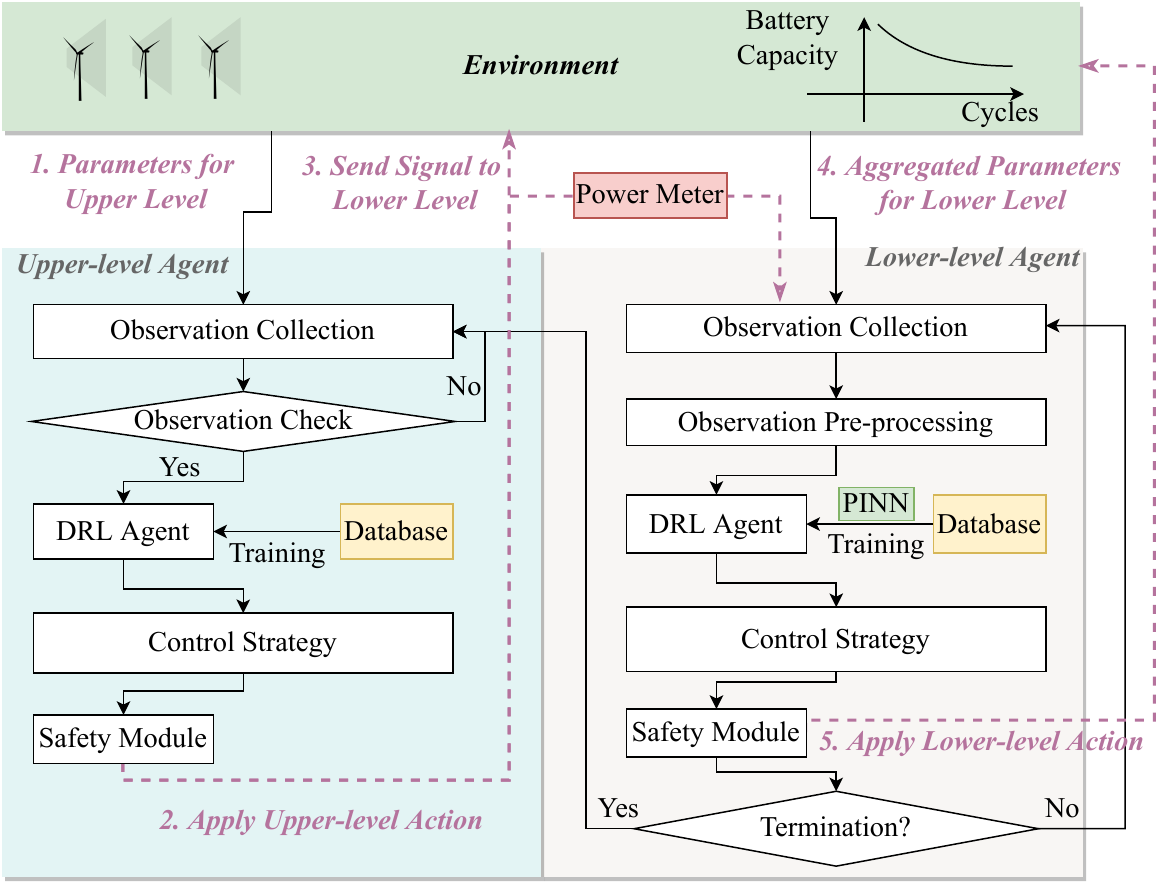}
    \caption{The coordinated bi-level control framework.} 
    \label{f2}
\end{figure}

To solve the aforementioned problem, a coordinated bi-level WSIS PSC framework is proposed, as depicted in Fig. \ref{f2}. The following steps outline the detailed procedure. First, the upper-level controller gathers observations from the environment, including the free stream speed and direction measured by the anemometer, as well as the on-grid price for wind power. Subsequently, the free stream velocity will be preprocessed to a suitable threshold to satisfy the cut-in and cut-off speed. If the free stream velocity surpasses the predefined threshold, the wind turbine will automatically shut down, and the upper-level controller will wait until the next decision time to reassess the observations and repeat the steps. When the system observes the current observation and the accumulative reward during the $f$ minutes, data are saved to a database to train the RL agent. The agent then calculates the real-time control strategy using the agent’s policy. The action should be checked through a safety module in case some improper operations damage the turbine. Furthermore, the upper-level action is conveyed to the lower-level controller as a guiding signal, which indicates the amount of wind power to allocate.

Based on the current state of the system, the lower-level module provides the control strategy with a higher frequency, precisely at a one-minute period. After aggregating the observations from both the environment and the upper-level signal from the power meter, the observations undergo normalization to an appropriate scale. The training process of the RL agent is similar to the upper-level controller, with the addition of a physics-informed neural network to expedite training. The strategy derived from the upper-level agent undergoes testing through a safety module to prevent battery charging/discharging actions from exceeding the maximum charging/discharging power limits. It's noteworthy that the upper-level controller generates actions every $f$ minutes, while the lower-level controller operates within the control loop until it reaches the maximum time steps, at which point the lower-level controller terminates the current control loop.

Due to the stochastic nature of wind, the sequential decision-making involved, and the adherence to the Markov property, the above-mentioned control problem can be reformulated as a Markov Decision Process (MDP). Customize distinct agents for each level within this coordinated control framework.

\subsubsection{Upper-level Agent} The upper-level agent is defined as follows.

\textit{State}: The upper-level agent controls the multiple wind turbines in the wind farm. Considering the target of the system, the state includes the incoming free-stream wind speed $U_{inf}^t$, wind direction $\varphi^t$, and electricity price $Pr^t$ in time $t$, i.e., $s^t_U=(U_{inf}^t, \varphi^t, Pr^t)$. Note that the free stream wind speed $U_{inf}^t$ should satisfy the Eq. (\ref{cut-wind}).
If the wind speed $U_{inf}^t$ falls outside the range, the state is invalid.

\textit{Action}: The action is defined as the axial induction factors of the turbines, which can be expressed as a vector: $a_U^t=\left(\alpha_1^t,\alpha_2^t,\ldots,\alpha_n^t\right)$. 
To ensure the safety, rather than directly applying $\alpha_i^t (i\in \mathcal{N}_\Psi)$ to the environment, a check model is utilized. If the action is infeasible, meaning $\alpha_i^t < 0$ or $\alpha_i^t > \frac{1}{2}$ for some $i\in \mathcal{N}_\Psi$, the agent is required to adjust the action to 
$\widetilde{a_U^t}=\left(\widetilde{\alpha_1^t},\widetilde{\alpha_2^t},\ldots,\widetilde{\alpha_n^t}\right)$ 
according to the following rule.
\begin{equation}
\widetilde{\alpha_i^t}=
    \left\{
    \begin{aligned}
    &\frac{1}{2},     &      & {\alpha}_i^t>\frac{1}{2}\\
    &\alpha_i^t,       &      & 0\le\alpha_i^t\le\frac{1}{2}\\
    &0,     &      & {\alpha}_i^t<0\\
    \end{aligned} \right.
\end{equation}

\textit{Reward}: The target of the upper-level agent is to maximize the wind farm's total profit and keep the action within the safe range. A normal way to deal with constraint is to use the violation as a punishment in the reward function. Therefore, the reward function consists of two parts: the first is the total power generation of the wind turbines, while the second is the penalized term for unsafe actions. The cumulative reward $r_U^t$ in a control interval will be received after $f$ minutes:
\begin{equation}
    r_U^t= \sum_{t^\prime=t}^{t+f-1} ({Pr^{t^\prime}\cdot \sum_{i\in \mathcal{N}_\Psi} P_i^{t^\prime}}) - G_U
\end{equation}
\begin{equation}    
G_U=\kappa\cdot \lVert a_U^t-\widetilde{a_U^t} \rVert
\end{equation}
where $P_i^{t^\prime}$, $U_i^t$, and $\varphi_i^t$ are the power generation, incoming wind speed, and wind direction of the turbine $i$ in time $t^\prime$, respectively. $\kappa$ is the penalized coefficient and $\lVert \cdot\rVert$ denotes the Manhattan distance. 

\subsubsection{Lower-level Agent}
The lower-level agent is defined as follows.

\textit{State}: The lower-level agent allocates the wind power generation based on information including historical power to grid. After the upper-level controller optimizes the turbines, the lower-level agent chooses the action according to the current environment state $s_U^t$, the grid-connected power of the previous time $P^{t-1}_G$, and the signal $g^t$ from the upper level. To smooth the power output to the grid, the upper-level signal $g^t$ is addressed as the total power generation $P_W^t$ for each time step. Although the induction factor $\widetilde{a_U^t}$ remains unchanged, $P_W^t$ still varies as a result of the changing wind velocity. Accordingly, the state is defined as: $s_L^t=\left[s_U^t, P^{t-1}_G, P_W^t\right]$.

\textit{Action}: The lower-level action $a_L^t$ is defined as the quantity of the charging or discharging power at time step $t$, defined as a continuous variable $a_L^t = P_B^t$ constrained in Eq. (\ref{BESS_2}), which ultimately determines the power flow to the grid by Eq. (\ref{flow_eq}). Invalid actions will be clipped to the boundary value during the learning process and punished according to the degree of the violation. 

\textit{Reward}: To mitigate the power fluctuation and maximize the total profit, the effectiveness of grid-connected power smoothing, the reduction in degradation and electricity purchasing costs, and the safety of the power flow are taken as the reward for the lower-level agent. By incorporating the safety constraint as a penalization term $G_L$, the reward function is meticulously crafted to summate three items, where two specific coefficients $\beta_i$ $(i=1,2)$ are assigned to delineate their relative significance.
\begin{equation}
    r_L^t = - [C_{deg}^t+ Pr^t\cdot(P_B^t-P_W^t)]
    -\beta_1\cdot P^t_{VG} - \beta_2\cdot G_L   
\end{equation}
\begin{equation}   
    G_L= |P^t_B+P_{max}^{dis}|+|P^t_B-P_{max}^{ch}|-P_{max}^{ch}-P_{max}^{dis}
\end{equation}

\subsection{MA-DDPG Algorithm }
To tackle the complexity and information flow of the given bi-level MDP problem, multi-agent DDPG (MA-DDPG) is specially developed to learn the optimal policy. Unlike DDPG, the centralized fully cooperative MA-DDPG algorithm comprises several actors (a policy network $\mu(s|\theta^\mu)$ with $\theta^\mu{}$ as the parameter used to approximate the deterministic policy function) which are used to execute different tasks according to the local observation. 
Each actor has a corresponding critic (a value network $Q(s,a|\theta^Q)$ with $\theta^Q$ as the parameter used to simulate the action-value function) that is used to guide the actions of the actors. 

For the coordinated control problem in WSIS, MA-DDPG employs a centralized training and decentralized execution architecture. The critics collectively share both actions and observations across all agents, while each actor independently acts based on its state. The actor’s parameter $\theta^\mu_k$ and the critic’s parameter $\theta^Q_k$ of agent $k$ $(k \in \{U,L\})$ are updated by:
\begin{equation}
\label{grad0}
    \mathcal{L}\left(\theta^Q_k\right) = \mathbb{E}\left[y^t - Q_k\left(s^t,a^t|\theta^Q_k\right)\right]^2 
\end{equation}
\begin{equation}
\label{yi}
    y^t = r^t_k+\gamma\max\limits_{a^t_k}{Q^\prime_k\left(s^{t+1},a^{t+1}\middle|\theta_k^{Q^\prime}\right)}
\end{equation}
\begin{equation}
\nabla_{\theta^\mu_k}J_k=\mathbb{E}\left[\nabla_{a_k}Q_k\left(s,a|\theta^Q_k\right)\middle|_{a_k=\mu_k\left(s_k\right)}\nabla_{\theta^\mu_k}\mu_k(s|\theta^\mu_k)\right]
\label{grad1}
\end{equation}
where $\mathcal{L}$ is the loss function for value network, $\mathbb{E}$ is the expectation of samples in random batch, and $\gamma\left(0<\gamma<1\right)$ is the discounted factor. $a_k$, 
$s_k$, $\mu_k$, $Q_k$ and $J_k$ are the action, state, policy network, value network and the cumulative discounted reward of the agent $k$, respectively. $\mu_k^\prime(s|\theta^{\mu^\prime_k})$ and $Q_k^\prime(s,a|\theta^{Q^\prime_k})$ are target networks to slowly update the learned actor and critic network with identical initial parameters \cite{van2016deep}, which are improved by a soft updating process as follows:
\begin{equation}\label{soft_update}
\begin{split}
    \theta_k^{Q^\prime} \gets \tau \theta_k^{Q} +(1-\tau)\theta_k^{Q^\prime}\\
    \theta_k^{\mu^\prime} \gets \tau \theta_k^{\mu} +(1-\tau)\theta_k^{\mu^\prime}
\end{split}
\end{equation}
where $\tau$ is the hyper-parameter for the target network to update.

\begin{figure}[t!]
    \centering 
    \includegraphics[width = 0.6\linewidth]{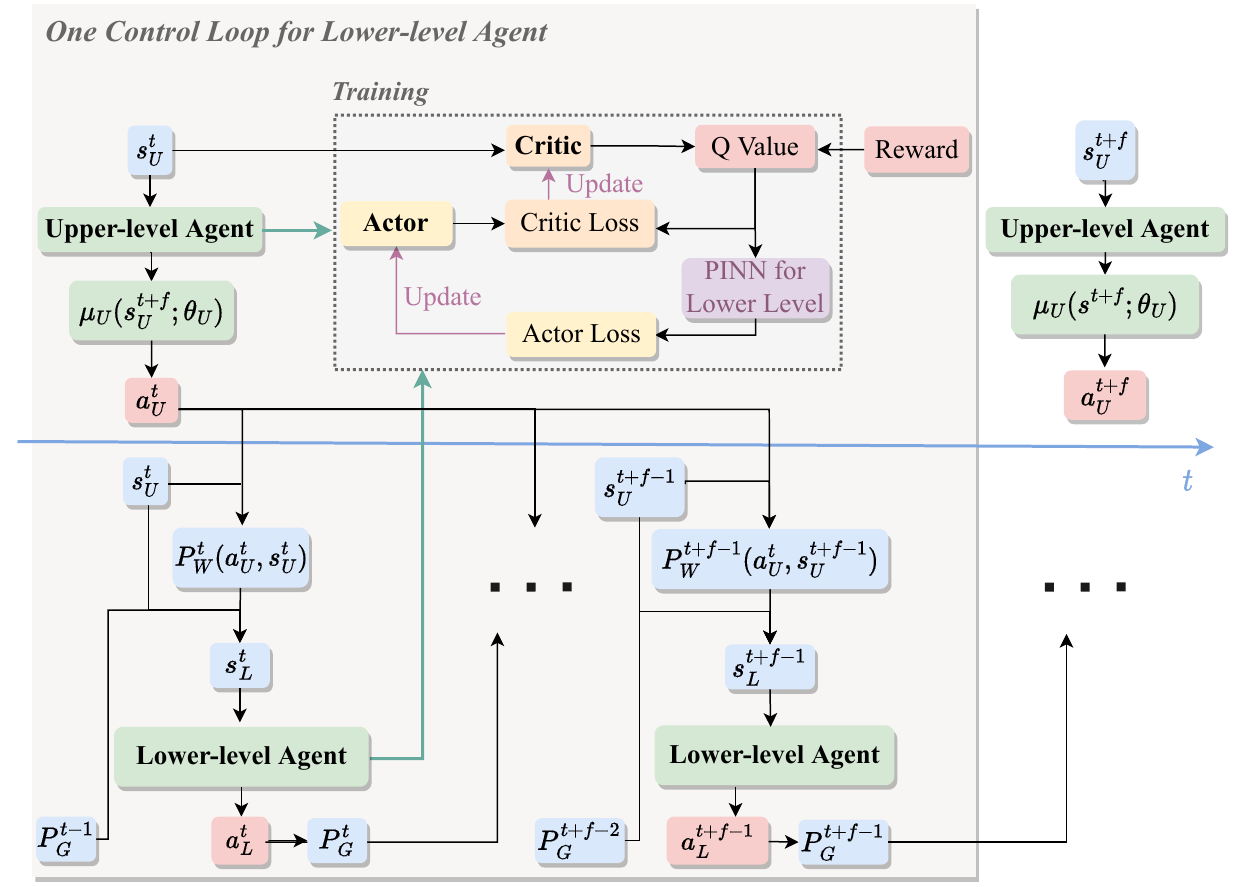}
    \caption{The workflow of the MA-DDPG. } 
    \label{f3}
\end{figure}

Specifically, as shown in Fig. \ref{f3}, the upper-level agent receives the raw states $s_U^t$ and produces its action $a_U^t$ by its actor network $\mu_U$. $a^t_U$ persists unchanged for the subsequent few time steps due to the decision frequency. For each state $s^{t^\prime}(t^\prime \in \{t,t+1,...,t+f-1\})$, $a^t_U$ corresponds to a signal $P^{t^\prime}_W$, which will be transmitted to the lower-level agent together with raw state $s_U^{t^\prime}$ and historical power output $P_G^{t-1}$. The lower-level agent integrates this information as its state $s_L^{t^\prime}$, producing its action $a_L^{t^\prime}$ using the actor network $\mu_L$ similarly and getting an instant reward $r_L^{t^\prime}$. 
The lower-level agent terminates either when the episode ends or when the maximum time step $f$ is reached. Subsequently, the upper-level agent obtains the cumulative reward $r_U^t$. The training process is similar for both agents using Eq. (\ref{grad0})-(\ref{grad1}) and their network parameters $\theta_U$ and $\theta_L$ are revised. Notably, PINN is designed for the lower-level training procedure, as elucidated in the subsequent section. Following this, the iterative process continues until each agent converges.

\subsection{Physics-informed Neural Network}

The proposed muti-agent DRL consists of two levels of policies and agents, thus, the interaction between different levels adds additional complexity and training time. Especially for the lower-level agent, its reward encompasses various facets, rendering the allocation of the optimal power flow a huge challenge. Therefore, considering the dynamics of power fluctuation for optimal policy, the PINN-based method is introduced to enhance the efficiency of the training process of the multi-agent DRL.

For the physical models of a general dynamic system, PINN incorporates its physical constraints into the neural network via differential equations involving time and other physical inputs. The general form \cite{raissi2019physics} can be expressed as:
\begin{equation}
\label{general PINN}
    \frac{\partial u(t,s)}{\partial t}+N(u)=0
\end{equation}
where $u(t,s)$ represents the potential solution, $s$ is the input state vector, and $N(\cdot)$ is an operator that can encapsulate a series of mathematical physics. 

When applied to this work, the solution is defined as the lower-level action $P_B^t=u(t,s_L)$ with lower-level state $s_L$. Assume that the power transferred to the grid is invariant to simulate the ideal dynamic system. Thus, the changes in power $P_{G}^t$ remain zero over time $t$. The invariable physical equation about the optimal power smoothing objective is derived as:
\begin{equation}
\label{pinn}
    \frac{\partial P_{G}^t}{\partial t} = 0
\end{equation}
where $P_{G}^t$ is identified as the variable related to the lower-level action $P_B^t$ by Eq. (\ref{flow_eq}). Accordingly, the above equation can be elegantly reformulated as a differential equation related to the solution $P_{B}^t$:
\begin{equation}
\label{pinn_WSIS}
     \frac{\partial P_{B}^t}{\partial t} + N(P_{B}^t)= 0
\end{equation}
which can be regarded as the special case of Eq. (\ref{general PINN}) for WSIS.
Define $f(t,s_L)$ to be given by the left-hand-side of Eq. (\ref{pinn_WSIS}), i.e.,
\begin{equation}
\label{fts}
    f(t,s_L) := \frac{\partial P_{B}^t}{\partial t} + N(P_{B}^t)
\end{equation}
and $u(t,s_L)$ has been approximated by policy network $\mu_L$. The policy network $\mu_L$ and physics-informed network $f(t,s_L)$ have the same parameters, while the activation functions are different due to the operator $N$. Therefore, the physics-informed network $f(t,s)$ can be utilized for the training of the lower-level agent. The shared parameters between the neural networks can be learned by minimizing the integrated mean squared error loss:
\begin{equation}
    \mathcal{L}_{PINN}=\omega_u\mathcal{L}_u+\omega_f\mathcal{L}_f
\end{equation}
where $\omega_u $ and $\omega_f $ are the weights to balance the interplay between the two loss terms. $\mathcal{L}_u$ and $\mathcal{L}_f$ are the data error and the physical information error of the neural network, respectively. In WSIS, $f$ represents the physical laws of power smoothness:
\begin{equation}
    \mathcal{L}_f=\frac{1}{N_f}\sum_{i=1}^{N_f}\left|f\left(t^i,s^i\right)\right|^2
\end{equation}
where $N_f$ is the batch size and $\{\left(t^i,s^i\right)\}$ are sets of experiences in the sampled batch. Hence the actor network is updated by the gradient ascent method:
\begin{equation}
    \nabla_{\theta^\mu}J \approx E_s[\omega_u\cdot\left(\nabla_aQ\left(s,a\middle|\theta^Q\right)|_{a=\mu\left(s_i\right)}\nabla_{\theta^\mu}\mu\left(s\middle|\theta^\mu\right)\right)+\omega_f\cdot\left|f\left(t,s\right)\right|^2]    
\label{grad2}
\end{equation}
The output of PINN and the associated physics-based information of environment states are provided as components to the RL agent’s neural network loss function during its training. It is noteworthy that, since $P_{G}^t$ has already been calculated for state derivation, there is no need for additional storage or computation of $P_{B}^t$ and $N(P_{B}^t)$ in $f(t,s)$. 
Considering physical laws during model training constrains the range of feasible solutions for neural network parameters, consequently reducing the demand for extensive training data and the size of the neural network. 
 The detailed process of PINN-assisted Multi-agent DDPG (PAMA-DDPG) is shown in Algorithm \ref{alg1}.

\begin{algorithm}[htpb]
    \caption{PINN-Assisted Multi-Agent DDPG}
    \begin{algorithmic}[1]
        \renewcommand{\algorithmicrequire}{\textbf{Input:}}
        \renewcommand{\algorithmicensure}{\textbf{Output:}}
        \REQUIRE initial  Q-function parameters ${\{\theta}_L^Q,\theta_U^Q\}$, policy parameters ${\{\theta}_L^\mu,{\theta}_U^\mu\}$, empty replay buffer $\{{RB}_L, {RB}_U\}$
        \ENSURE network parameters
        \STATE Set target parameters $\{{\theta}_L^{Q\prime},\theta_U^{Q\prime}\} \gets \{{\theta}_L^Q,\theta_U^Q\}$ and $\{{\theta}_L^{\mu\prime},\theta_U^{\mu\prime}\} \gets \{{\theta}_L^\mu,\theta_U^\mu\}$   
            \FOR {$episode=1, num\_episodes$}
            \STATE Initialize environment and get initial state $s^t$;
            \STATE Initialize random process $N_U^t,N_L^t$ for action exploration
            \WHILE{$s$ is not terminal}
                \STATE Observe $s_U^t$, $0\gets R_U$
                \STATE Select upper-level action  $a_U^t\gets\mu_U\left(s_U^t|\theta_U^\mu\right)+N_U^t$
                \STATE Observe $s_L^t\gets\{s_U^t,P_G^{t-1}, P_W^t\}$
                \WHILE{not ($s$ is terminal or lower level terminates)}
                    \STATE Select lower-level action $a_L^t\gets\mu_L\left(s_L^t|\theta_L^\mu,P_W^t\right)+N_L^t$
                    \STATE Observe $s_U^{t+1}$ and $s_L^{t+1}\gets\{s_U^{t+1}, P_G^{t}, P_W^{t+1}\}$
                    \STATE Obtain instant reward $r_H^t$ and $r_L^t$ for two levels
                    \STATE Store transition\ $(s_L^t,\ a_L^t,\ r_L^t,\ s_L^{t+1})$ in ${RB}_L$
                    \STATE Sample mini-batch from $RB_L$ and update parameters $\theta_L$ of the lower level by (\ref{grad0}), (\ref{yi}), (\ref{grad2}) and (\ref{soft_update})
                    \STATE $R_U\gets R_U+r_U^t$, $s^t\gets s^{t+1}$, $s_L^t\gets s_L^{t+1}$
                \ENDWHILE
                \STATE Store transition $(s_U^t,\ a_U^t,\ R_U^t,\ s_U^{t+1})$ in ${RB}_U$
                \STATE Sample mini-batch from $RB_U$ and update parameters $\theta_L$ of the lower level by (\ref{grad0}), (\ref{yi}), (\ref{grad1}) and (\ref{soft_update})
            \ENDWHILE
            \ENDFOR 
    \end{algorithmic}
    \label{alg1}
\end{algorithm}

\section{Experimental Results}
\label{Section IV}
\subsection{Setup}
\begin{figure}[t!]
    \centering 
    \includegraphics[width = 0.6\linewidth]{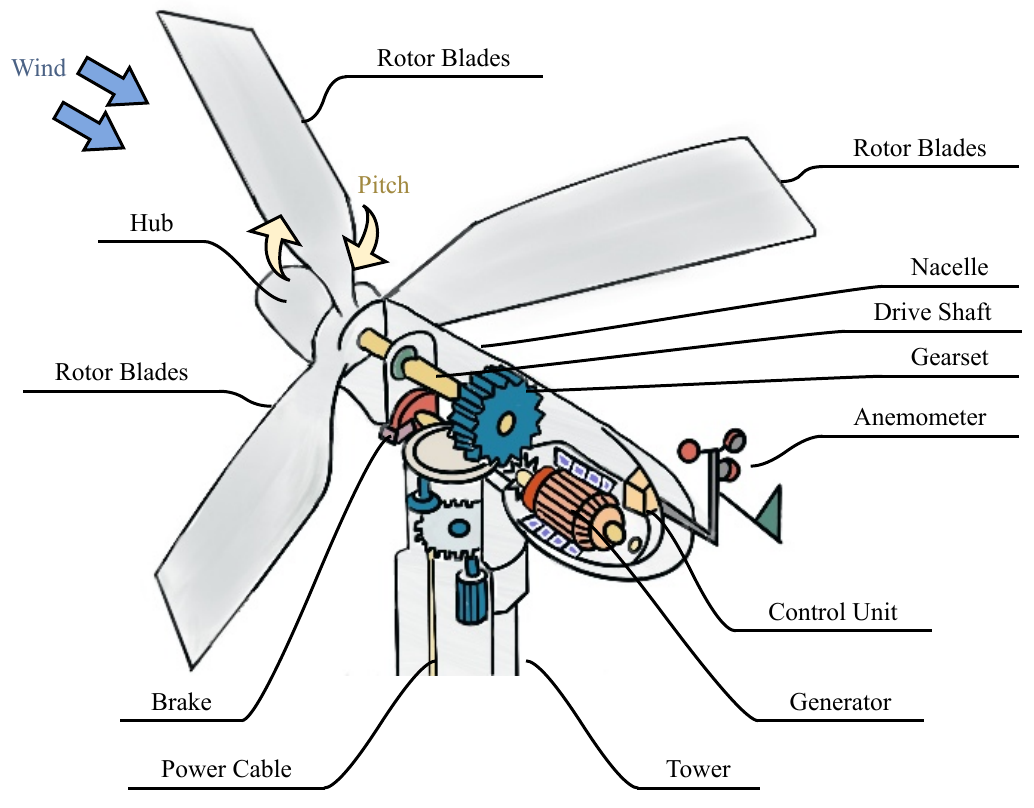}
    \caption{The wind turbine structural system.} 
    \label{f_turbine}
\end{figure}

To evaluate the effectiveness of the proposed PAMA-DDPG algorithm, we employ the dynamic control-oriented wind farm simulator, WindFarmSimulator (WFSim) \cite{boersma2016control}, for conducting comprehensive case studies. WFSim is a dynamic medium fidelity control-oriented model that predicts the flow velocity vectors in a wind farm using the spatially and temporally discretized 2D Navier-Stokes equations, which strike a balance between simulation fidelity and computational complexity. In practical applications, the wind turbine with three rotor blades is typically positioned upwind of the tower and the nacelle, as depicted in Fig. \ref{f_turbine}. The optimal induction factor, determined by the centralized control system, is then conveyed to the internal control unit of the turbine to regulate the pitch and the rotor speed.

To better restore the real world, real-world wind speed data from Alternative Energy Development Board (AEDB) \cite{winddata} is incorporated into the simulations, with 10-minute average values being interpolated to minute-by-minute values.  
For the purpose of our study, it is assumed that the direction of freestream velocity remains fixed in the linear topology. In accordance with the renewable energy feed-in tariff policy, the on-grid price is maintained as a constant value \cite{ma2011grid}. As per the specifications for wind farms' active power regulation and control \cite{apcchina}, the fluctuation threshold is defined as 3 MW/min. For BESS, the lithium-iron battery is utilized with energy conversion efficiencies $\eta_{ch}=\eta_{dis}=0.98$ \cite{barra2021review}. The data for battery degradation calculation comes from a real-world application in Rwanda \cite{bordin2017linear}. Complete details of the parameters for WFSim, BESS, PARK, and DRL can be found in Table \ref{para}. The wind farm simulation is conducted using Matlab R2020a, and the algorithms are trained on a PC with an Intel Core i9 processor and an NVIDIA RTX 2060 GPU.

\begin{table}[htpb]\scriptsize
\setstretch{1.2}
  \centering
  \caption{Hyperparameters Used In Experiments}
    \begin{tabularx}{0.6\linewidth}{lYYYY}
    \toprule
    \specialrule{0em}{0.3pt}{0.5pt}
    \toprule
    \textit{Parameter} & \textit{Value}   \\
    \midrule
    \specialrule{0em}{0pt}{0pt}
    \rowcolor{gray} Powerscale & 0.95 \\
                    Forescale & 1.5  \\
    \rowcolor{gray} Wake Expansion Coefficient & 0.08  \\ 
                    Air Density ($\rm kg/m^3$) & 1.2  \\ 
    \rowcolor{gray} Turbine Diameter ($\rm m$) & 100  \\ 
                    Battery Capacity ($\rm MWh$) & 6  \\
    \rowcolor{gray}   $RE$ (\%) & 89  \\
                    $C_R$  (\$) & 900\\
    \rowcolor{gray} $L_T$ ($\rm kWh$) & 1344 \\
                    Electricity Price ($\rm \$/MWh$) & 300  \\ 
    \rowcolor{gray} Energy Conversion Efficiency & 0.98 \\
                    $\kappa$  & 10 \\
    \rowcolor{gray} $\beta_1$ &5 \\
                    $\beta_2$  & 10 \\
    \rowcolor{gray}  Batch Size & 32  \\
                    Critic-Network Learning Rate & 0.01  \\
   \rowcolor{gray} Actor-Network Learning Rate & 0.0001  \\ 
                    Target Network Hyper-Parameter & 0.001  \\ 
    \specialrule{0em}{0pt}{0pt}
    \bottomrule
    \specialrule{0em}{0.5pt}{0.3pt}
    \bottomrule
    \end{tabularx}
  \label{para}
\end{table}

Prior to delving into the assessment of the proposed PAMA-DDPG approach, it is imperative to establish a framework comprising several wind farm control methodologies, with the intention of facilitating thorough comparison and validation.

\textit{1) Model Predictive Control (MPC)}: At each time step, the MPC method forecasts the forthcoming free-stream velocity, guiding the analytical Jensen Park model with ADM to estimate the wake effect and the power generation of each wind turbine. Subsequently, an optimization model is formulated to derive the BESS schedules – establishing a benchmark.

\textit{2) DDPG}: Traditional single-agent DDPG method can solve complex MDPs but cannot be directly applied to deal with different control frequencies. Therefore, an agent is designed to control both the turbines and the batteries simultaneously with a reduced control frequency, which is the control frequency of the wind turbine in this case. Suppose BESS has the same control frequency as the turbines and the DDPG agent outputs a two-level action every five minutes.
For a fair comparison, we use the same network architectures as those for PAMA-DDPG. The actor and critic networks are composed of one input layer, three hidden layers, and one output layer, respectively. The neuron number in each hidden layer is set as (400, 300, 400). The Rectified Linear Unit is used as the activation function.

\textit{3) MA-DDPG}: This method has the identical multi-agent architecture with the proposed PAMA-DDPG except for the PINN structure, which aims to illustrate superior learning performance when integrating physics information.

The objective of the predefined WSIS control problem is to maximize the total profit of the WSIS as well as mitigate power fluctuation. To judge the overall performance of a given control policy, we need to consider both the profit and the fluctuation. The total profit is the power generation profit minus the battery degradation cost, as defined in Eq. (\ref{obj}). More specifically, we consider the power fluctuation in two metrics. Define the fluctuation severity of the power output $FS$:
\begin{equation}
    FS=\sum_{t=1}^{N}P^t_{FG}
\end{equation}
where $N$ denotes the length of a testing wind sequence. Define violation occurrence $VO$ of the power output as the times that WSIS’s power output exceeds the prescribed threshold:
\begin{equation}
VO=\sum_{t=1}^{N}{max(sign(}P^t_{FG}-P_{FG, max}),0)
\end{equation}
where the function $sign\left(\cdot\right)$ is utilized to indicate excessive fluctuation. The two different metrics are used to quantify the severity and occurrence of violations or deviations from the stable power output. 

\begin{figure}[htpb]
    \centering 
    \includegraphics[width = 1\linewidth]{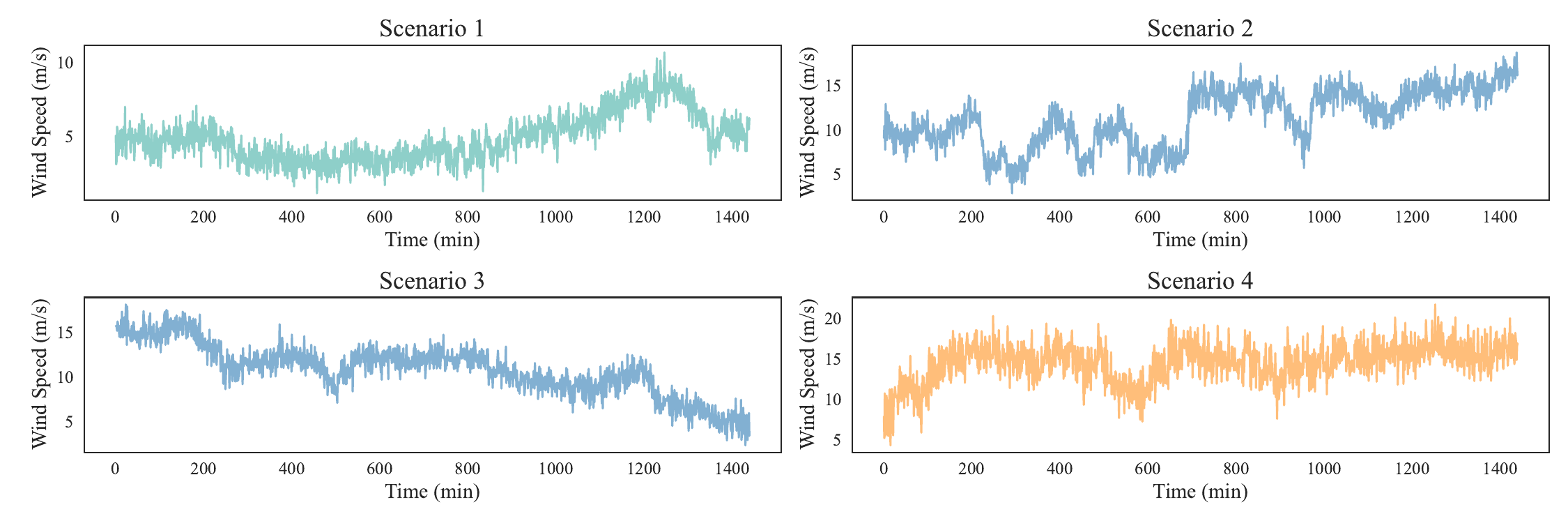}
    \caption{One-day wind speed profile for 4 testing scenarios.} 
    \label{f4}
\end{figure}

Moreover, in order to mitigate the impact of the ever-changing wind velocity, four typical one-day wind sequences are employed as testing scenarios, thereby showcasing the algorithms' adaptability to varying wind patterns, which is showcased in Fig. \ref{f4}. Scenario 1 exemplifies conditions of low wind velocity, while scenarios 2 and 3 depict moderate wind speeds, representing the most prevalent circumstances. Scenario 4, on the other hand, illustrates a state of high wind velocity. For better illustration, the average metrics over the four scenarios are utilized to compare the general performance.

\subsection{Optimal Results Comparison}
 To provide a clear demonstration of the optimal results achieved by the proposed algorithm, MPC’s performance is utilized as the baseline, and the relative metrics of other methods are calculated. By comparing the performance of MPC, DDPG, MA-DDPG, and PAMA-DDPG algorithms, we can effectively assess the effectiveness and superiority of the proposed approach.
 \begin{figure}[htpb]
    \centering 
    \includegraphics[width = 0.6\linewidth]{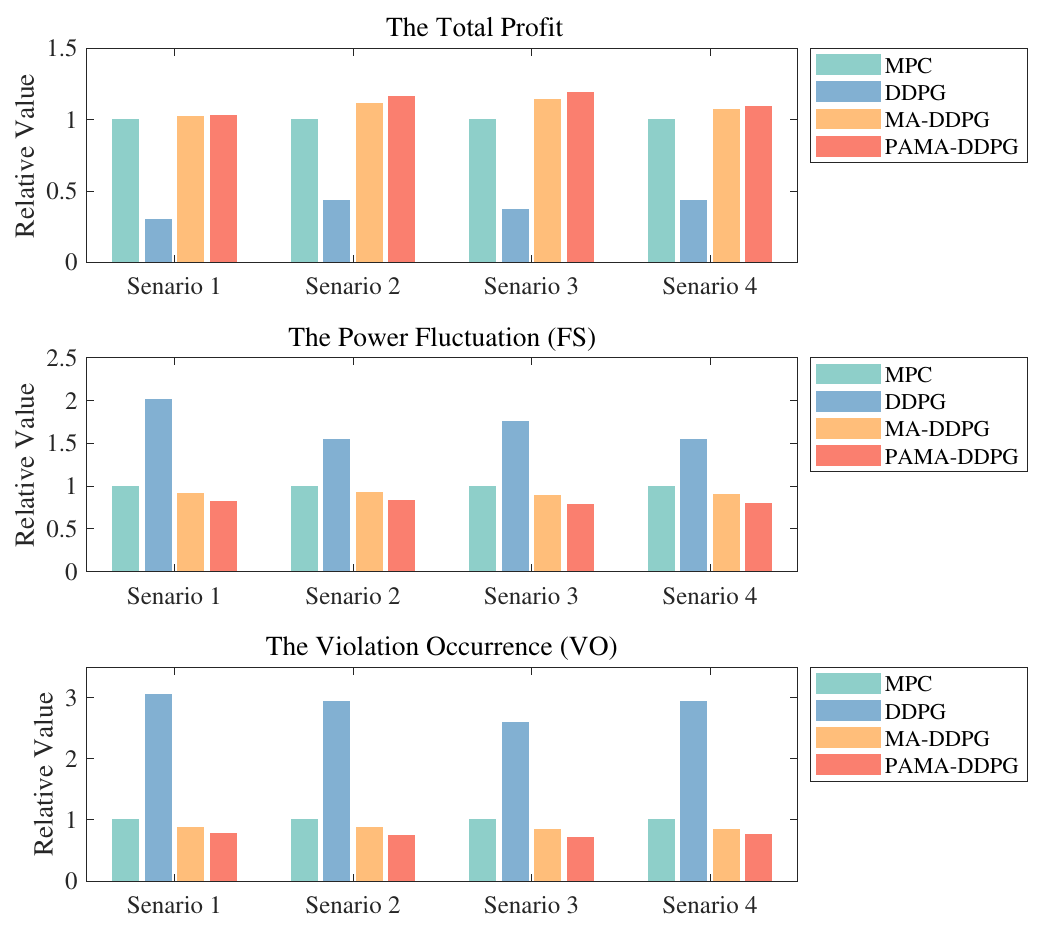}
    \caption{Relative metrics of 4 methods in various scenarios.} 
    \label{f5}
\end{figure}

Fig. \ref{f5} exhibits that the proposed PAMA-DDPG algorithm can always reach the maximum total profit and minimum fluctuation in the WFSim environment. On average, PAMA-DDPG demonstrates improvements of roughly +11\%, -19\%, and -22\% in total profit, $FS$, and $VO$ compared to the MPC method. The base values for MPC are $ \$840.55, 19.01$MW, and $23$ for total profit, $FS$, and $VO$ respectively. 
As the single-agent DDPG control operates at a reduced frequency, it yields unsatisfactory performance. The reason is that the agent cannot react to the environmental change in time.
These comparative results underscore the ability of both PAMA-DDPG and MA-DDPG approaches to acquire the optimal policy across all scenarios. This successful acquisition of the optimal policy by the bi-level structure serves as evidence of the effectiveness and robustness of the proposed algorithm. Conversely, the conventional DDPG algorithm proves inadequate in addressing such intricate challenges, consequently impeding its ability to acquire a proficient policy.  

It is noteworthy that while two fluctuation-related metrics demonstrate notable improvement, the enhancement in total profit is comparatively modest. This discrepancy arises due to the substantial MPC baseline and the inevitable trade-off that entails sacrificing some power output in order to achieve a smoother power profile. Specifically, it is inevitable to either charge or discharge the battery to stabilize the power injection into the grid, consequently incurring degradation costs.

\begin{figure}[b!]
    \centering 
    \includegraphics[width = 0.6\linewidth]{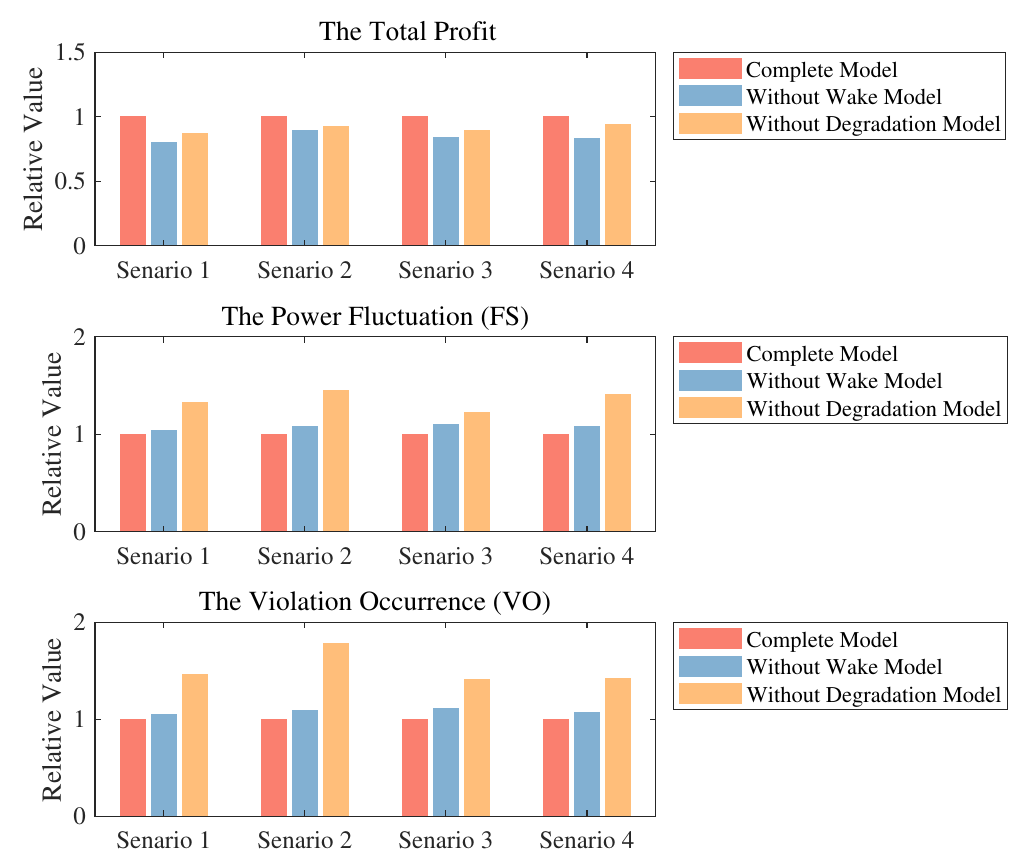}
    \caption{Relative metrics of different environment model completeness in various scenarios.} 
    \label{f6}
\end{figure}
\subsection{Verification of Model Completeness}
Within the proposed framework, we adopt a comprehensive approach by considering the influence of both the wind wake model and the battery degradation model. In order to evaluate the impact of model completeness, two controlled experiments are designed to compare the proposed control strategies.

Firstly, if the wake model is not considered, the upper layer control problem could have an analytical solution. By utilizing the power generation formula of the wind turbine, since the induction factor $A$ is constrained by 0 and 1/2, the maximum power extraction $P^\ast$ for the single turbine is determined as $A^\ast = cos(\varphi_y)/3$, while the coefficient of power, denoted as $C_p^\ast$, can be expressed as $16\cos^3\ (\varphi_y)/27$. Secondly, the battery degradation model is directly related to the learning goal for the lower-level controller. Without considering the degradation, the performance feedback obtained from control algorithms may be inaccurate, leading to unreliable operation of the system. To validate it, we test the proposed PAMA-DDPG algorithm in the four testing scenarios with varying model settings.

The results depicted in Fig. \ref{f6} highlight the positive influence of accurate and comprehensive models on the WSIS system's control objectives, particularly in maximizing total profit and minimizing power fluctuation/violation. Specifically, define the complete model as the baseline, for the model without wake effect, the average total profit, $FS$, and $VO$ are roughly 85\%, 107\%, and 108\%, respectively. For the model without battery degradation, the average total profit, $FS$, and $VO$ are 91\%, 135\%, and 152\%. By incorporating a complete environmental model, the PAMA-DDPG algorithm is empowered to make more informed decisions, leading to improved control performance.

\subsection{Verification of the Efficiency of PINN}
\begin{figure}[htbp]
    \centering 
    \includegraphics[width = 1\linewidth]{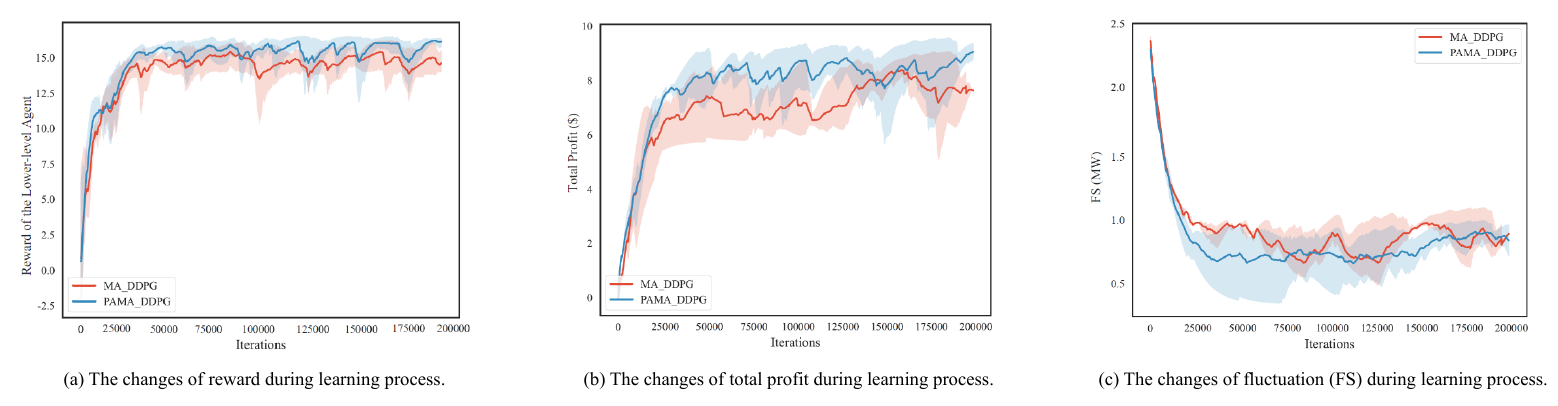}
    \caption{Learning process of the proposed approach.} 
    \label{f7}
\end{figure}

The lower-level controller utilizes the PINN during the training process. In this test, we compare the changes of rewards, total profit, and fluctuation ($FS$) against iteration times between the MA-DDPG and PAMA-DDPG algorithms to validate the efficacy of PINN. Due to the significant variability in wind velocity, the direct output remains highly volatile even with a fixed policy during the training process. To quickly visualize the policy modifications at each iteration, a fixed short-term five-minute wind speed sequence is utilized for testing, which is (8.74, 7.32, 4.50, 10.39, 6.66) m/s. To eliminate the randomness, each of the algorithms runs 10 times to draw the average learning performance (solid line) with standard deviation (shaded region) as shown in Fig.\ref{f7}, which illustrates the first 2e5 iterations. The average time for the two algorithms to choose an action and update the policy is 1e-8 and 1e-2 seconds, respectively.
In general, PAMA-DDPG and MA-DDPG learn the optimal policy within approximately 4e4 and 5e4 iterations, respectively.  It is clear in Fig. \ref{f7}(a) that the blue line rises faster than the red line before convergence, which means PINN can accelerate the learning efficiency. Furthermore, PAMA-DDPG also converges to a higher reward, greater profit, and reduced fluctuation, as shown in subplots of Fig. \ref{f7}. 
Specifically, PAMA-DDPG achieves a reward of 15.57, a profit of 8.18\$, and a fluctuation of 0.71MW at iteration 4e4, while MA-DDPG only obtains a reward of 14.69, a profit of 7.49\$, and a fluctuation of 0.96MW at iteration 5e4.
The inherited physics information in the PAMA-DDPG enhances its effectiveness for the specific WSIS control problem. 

\subsection{Robustness Analysis}

Given that the lower level is responsible for degradation cost and power fluctuation, the penalty factor $\beta_1$ plays a critical role in PAMA-DDPG, which balances their respective significance in the control task. Take five $\beta_1$ values in $[0,0.5,1,5,10]$, and test them in the four test scenarios. The average experimental results are quite stable in a wide range except 0, as shown in Table. \ref{hyper1}. When $\beta_1=0$, the influence of power fluctuation is disregarded, and the degradation cost becomes negligible, resulting in the highest levels of the three metrics. As $\beta_1$ becomes positive, both $FS$ and $VO$ experience a sharp decline, albeit with some sacrifice to total profit. 

\begin{table}[htpb]\scriptsize
\setstretch{1.2}
  \centering
  \caption{Performance of PAMA-DDPG for different $\beta_1$.}
    \begin{tabularx}{0.6\linewidth}{lYYYY}
    \toprule
    \specialrule{0em}{0.3pt}{0.5pt}
    \toprule
    \textit{$\beta_1$} & \textit{Total Profit}   & \textit{FS}  & \textit{VO}\\
    \midrule
    \specialrule{0em}{0pt}{0pt}
    \rowcolor{gray} 0 &     945.11 &    34.96 &    47\\
                    0.5 &   936.18 &    16.93 &    19\\
    \rowcolor{gray} 1 &     932.45 &    14.28&     20\\ 
                    5 &     933.01 &    15.39&     18\\
    \rowcolor{gray} 10 &    930.12 &    14.11&     18\\
    \specialrule{0em}{0pt}{0pt}
    \bottomrule
    \specialrule{0em}{0.5pt}{0.3pt}
    \bottomrule
    \end{tabularx}
  \label{hyper1}
\end{table}

We further verify that PAMA-DDPG works effectively for different energy conversion efficiencies by changing $\eta_{ch}$ and $\eta_{dis}$. Table. \ref{hyper2} shows that PAMA-DDPG can effectively learn the optimal policies for various energy conversion efficiency. The change of $\eta_{ch}$ and $\eta_{dis}$ has minimal impact on overall profit. This is due to the fact that wind power generation predominantly contributes to the total profit, making the profit associated with energy loss relatively insignificant.

\begin{table}[htpb]\scriptsize
\setstretch{1.2}
  \centering
  \caption{Performance of PAMA-DDPG for different energy conversion efficiency.}
    \begin{tabularx}{0.6\linewidth}{lYYYY}
    \toprule
    \specialrule{0em}{0.3pt}{0.5pt}
    \toprule
    \textit{$\eta_{ch}$}  & \textit{$\eta_{dis}$} & \textit{Total Profit}   & \textit{FS}  & \textit{VO}\\
    \midrule
    \specialrule{0em}{0pt}{0pt}
    \rowcolor{gray} 0.95 &   0.95 &  932.67 &    15.27 &    18\\
                    0.98 &   0.98 &  933.01 &    15.39 &    18\\
    \rowcolor{gray} 1.00 &   1.00 &  936.45 &    16.36&     19\\
    \specialrule{0em}{0pt}{0pt}
    \bottomrule
    \specialrule{0em}{0.5pt}{0.3pt}
    \bottomrule
    \end{tabularx}
  \label{hyper2}
\end{table}

\subsection{Case Analysis of Lower-level Strategy}
\begin{figure}[b!]
    \centering 
    \includegraphics[width = 0.6\linewidth]{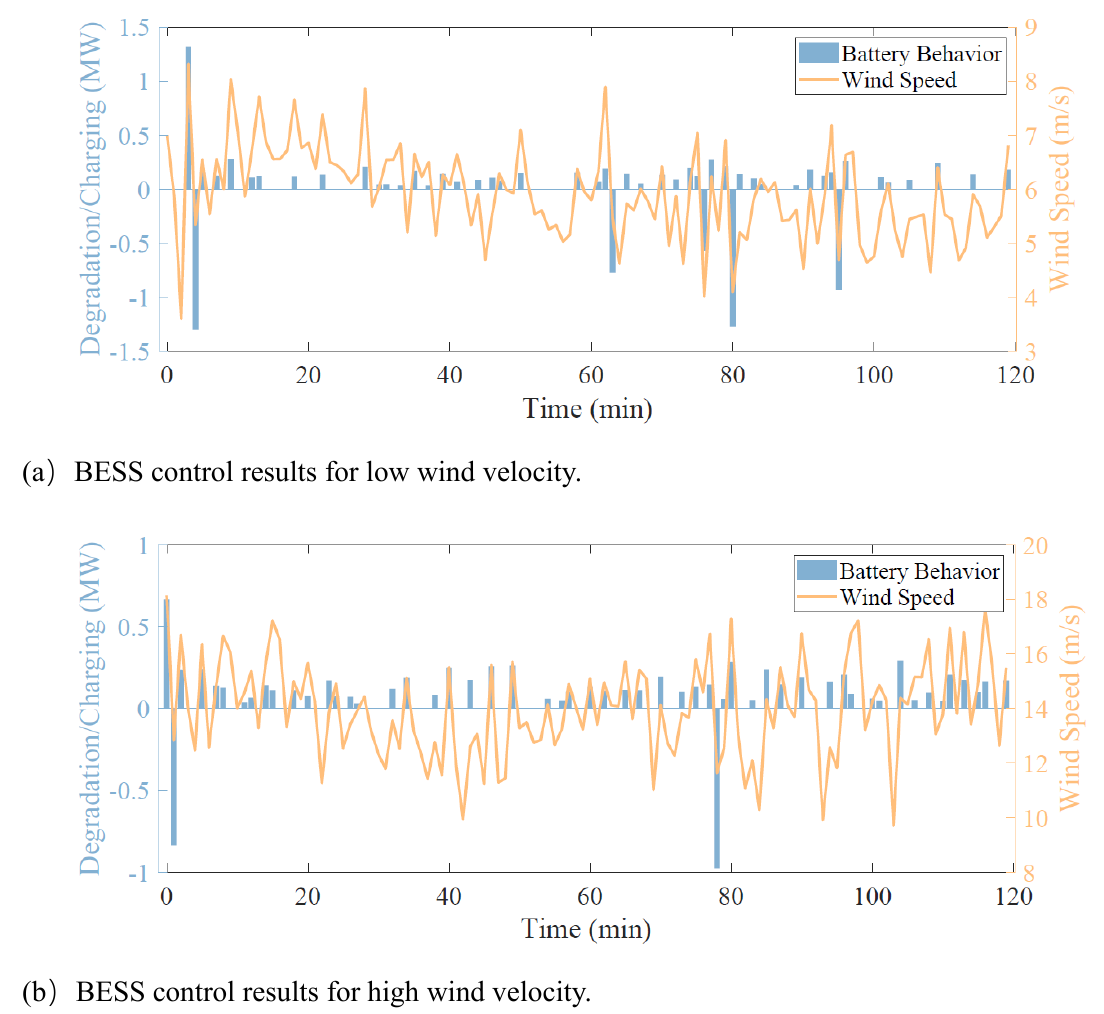}
    \caption{BESS control results obtained by the proposed PAMA-DDPG over 100 minutes.} 
    \label{f8}
\end{figure}

To visually demonstrate the effectiveness of the power smoothing strategy derived from the proposed algorithm, we analyze the charging/discharging behavior corresponding to wind speed fluctuations within a typical time interval. Two representative 120-minute intervals from the testing scenarios are selected, which exhibit pronounced fluctuations. Fig. \ref{f8} presents an intuitive illustration of how PAMA-DDPG successfully mitigates power variations over a 100-minute testing duration. Fig. \ref{f8}(a) and Fig. \ref{f8}(b) show the BESS behaviors for low wind velocity and high wind velocity, respectively. Consider the instance when time equals 4 min in Fig. \ref{f8}(a), with the elevation of wind speed from 3.61m/s to 8.32m/s, the BESS charges 1.32MW to retain the excess power. However, in the subsequent minute, as the wind speed descends to 5.33m/s, the BESS discharges 1.28MW to uphold stable power production. In most cases, PAMA-DDPG acquires a commendable charging strategy for the BESS as the wind speed ascends, while effectively discharging it as the speed diminishes. It's worth mentioning that BESS remains idle while the wind fluctuation remains within the accepted threshold to minimize degradation costs, as shown during time 50-60 in Fig. \ref{f8}(a). For the same reason, it is more likely to charge than to discharge, as charging behavior incurs degradation costs.
Combining two subfigures, BESS can accommodate both low and high wind velocities, and the extent of the charging/discharging behavior is mainly influenced by short-term wind fluctuations.
The experimental outcomes elucidate that the proposed PAMA-DDPG exhibits adaptability to diverse wind-changing patterns and possesses the capability to mitigate power output fluctuations.

\section{Conclusion}
\label{Section V}

This paper presented an innovative approach to address the PSC challenges in WSIS. A comprehensive coordinated control framework is developed considering the distinct control frequencies, wake effect, and battery degradation cost. To overcome the challenges of complex systems and environment randomness, a PAMA-DDPG algorithm was proposed, incorporating the power fluctuation differential equation into the loss function of the policy network. The feasibility and effectiveness of the proposed power smoothing control strategy have been demonstrated through extensive simulation scenarios using WFSim. The numerical results have verified the following.
\begin{itemize}
    \item The proposed PAMA-DDPG algorithm yielded an average total profit 11\% higher than that achieved by the MPC baseline. Furthermore, the average fluctuation and violation of the PAMA-DDPG algorithm were approximately 19\% and 22\% lower, respectively, compared to the MPC baseline.
    
    \item The learning process of PAMA-DDPG exhibited faster convergence compared to MA-DDPG. By leveraging PINN for loss design, convergence iterations for the algorithm were enhanced from 5e4 to 4e4. Specifically, PAMA-DDPG achieves a reward of 15.57 at iteration 4e4, while MA-DDPG only obtains a reward of 14.69 at iteration 5e4. The results of this study contribute to  PSC problems in WSIS, showcasing the potential of PIDRL. 
    
    \item The wake effect and battery degradation cost model contribute to a comprehensive framework formulation and result in better power output and power smoothing. The total profit declined about 9\%-15\%, while $FS$ and $VO$ increased about 7\%-35\% and 8\%-52\% respectively for deficient environmental models.
\end{itemize}

The proposed framework and algorithm offer valuable insights for optimizing WSIS total profit and improving the reliability and efficiency of renewable energy systems. 
Our potential future works will be focused on exploring the incorporation of other physical laws, thereby improving the algorithm's predictive accuracy and usefulness in practical applications. The incorporation of transfer learning and PINN is also a promising approach for expediting the learning process.
Moreover, the tricks of automatic hyperparameter tuning are also considered one of our future plans to streamline the process of selecting optimal hyperparameter configurations. Finally, models of electronic components and real-world experiments could be pursued to validate the findings and assess the practical implementation of the proposed approach in actual WSISs.

\section*{Acknowledgements}
\label{sec:Acknowledgements}
This work was supported in part by the Shenzhen Institute of Artificial Intelligence and Robotics for Society (AIRS); Shenzhen Key Lab of Crowd Intelligence Empowered Low-Carbon Energy Network (No. ZDSYS202206061 
 00601002); National Natural Science Foundation of China, Grant/Award Numbers: 72331009, 72171206.

\bibliographystyle{elsarticle-num} 
\bibliography{202308}

\begin{thebibliography}{10}
\expandafter\ifx\csname url\endcsname\relax
  \def\url#1{\texttt{#1}}\fi
\expandafter\ifx\csname urlprefix\endcsname\relax\def\urlprefix{URL }\fi
\expandafter\ifx\csname href\endcsname\relax
  \def\href#1#2{#2} \def\path#1{#1}\fi

\bibitem{yin2023multi}
L.~Yin, W.~Ding, Multi-objective high-dimensional multi-fractional-order optimization algorithm for multi-objective high-dimensional multi-fractional-order optimization controller parameters of doubly-fed induction generator-based wind turbines, Engineering Applications of Artificial Intelligence 126 (2023) 106929.

\bibitem{shivashankar2016mitigating}
S.~Shivashankar, S.~Mekhilef, H.~Mokhlis, M.~Karimi, Mitigating methods of power fluctuation of photovoltaic (pv) sources--a review, Renewable and Sustainable Energy Reviews 59 (2016) 1170--1184.

\bibitem{tang2018active}
X.~Tang, M.~Yin, C.~Shen, Y.~Xu, Z.~Y. Dong, Y.~Zou, Active power control of wind turbine generators via coordinated rotor speed and pitch angle regulation, IEEE Transactions on Sustainable Energy 10~(2) (2018) 822--832.

\bibitem{kim2016power}
Y.~Kim, M.~Kang, E.~Muljadi, J.-W. Park, Y.~C. Kang, Power smoothing of a variable-speed wind turbine generator in association with the rotor-speed-dependent gain, IEEE Transactions on Sustainable Energy 8~(3) (2016) 990--999.

\bibitem{uehara2011coordinated}
A.~Uehara, A.~Pratap, T.~Goya, T.~Senjyu, A.~Yona, N.~Urasaki, T.~Funabashi, A coordinated control method to smooth wind power fluctuations of a pmsg-based wecs, IEEE Transactions on energy conversion 26~(2) (2011) 550--558.

\bibitem{barra2021review}
P.~Barra, W.~De~Carvalho, T.~Menezes, R.~Fernandes, D.~Coury, A review on wind power smoothing using high-power energy storage systems, Renewable and Sustainable Energy Reviews 137 (2021) 110455.

\bibitem{lu2024advances}
P.~Lu, N.~Zhang, L.~Ye, E.~Du, C.~Kang, Advances in model predictive control for large-scale wind power integration in power systems: A comprehensive review, Advances in Applied Energy (2024) 100177.

\bibitem{zhang2020accommodate}
H.~Zhang, J.~Yang, X.~Ren, Q.~Wu, D.~Zhou, E.~Elahi, How to accommodate curtailed wind power: A comparative analysis between the us, germany, india and china, Energy Strategy Reviews 32 (2020) 100538.

\bibitem{wang2023coordinated_tse}
X.~Wang, J.~Zhou, B.~Qin, L.~Guo, Coordinated power smoothing control strategy of multi-wind turbines and energy storage systems in wind farm based on madrl, IEEE Transactions on Sustainable Energy (2023).

\bibitem{huang2019hierarchical}
S.~Huang, Q.~Wu, Y.~Guo, F.~Rong, Hierarchical active power control of dfig-based wind farm with distributed energy storage systems based on admm, IEEE Transactions on Sustainable Energy 11~(3) (2019) 1528--1538.

\bibitem{wang2023coordinated}
X.~Wang, J.~Zhou, B.~Qin, L.~Guo, Coordinated control of wind turbine and hybrid energy storage system based on multi-agent deep reinforcement learning for wind power smoothing, Journal of Energy Storage 57 (2023) 106297.

\bibitem{xiong2021optimal}
L.~Xiong, S.~Yang, S.~Huang, D.~He, P.~Li, M.~W. Khan, J.~Wang, Optimal allocation of energy storage system in dfig wind farms for frequency support considering wake effect, IEEE Transactions on Power Systems 37~(3) (2021) 2097--2112.

\bibitem{ahmad2016model}
M.~A. Ahmad, M.~R. Hao, R.~M. T.~R. Ismail, A.~N.~K. Nasir, Model-free wind farm control based on random search, in: 2016 IEEE international conference on automatic control and intelligent systems (I2CACIS), IEEE, 2016, pp. 131--134.

\bibitem{dong2020model}
H.~Dong, M.~Edrah, X.~Zhao, M.~Collu, X.~Xu, K.~Abhinav, Z.~Lin, Model-free semi-active structural control of floating wind turbines, in: 2020 Chinese Automation Congress (CAC), IEEE, 2020, pp. 4216--4220.

\bibitem{sutton2018reinforcement}
R.~S. Sutton, A.~G. Barto, Reinforcement learning: An introduction, MIT press, 2018.

\bibitem{cao2024survey}
Y.~Cao, H.~Zhao, Y.~Cheng, T.~Shu, G.~Liu, G.~Liang, J.~Zhao, Y.~Li, Survey on large language model-enhanced reinforcement learning: Concept, taxonomy, and methods, arXiv preprint arXiv:2404.00282 (2024).

\bibitem{zhao2020cooperative}
H.~Zhao, J.~Zhao, J.~Qiu, G.~Liang, Z.~Y. Dong, Cooperative wind farm control with deep reinforcement learning and knowledge-assisted learning, IEEE Transactions on Industrial Informatics 16~(11) (2020) 6912--6921.

\bibitem{zhu2022optimal}
J.~Zhu, W.~Hu, X.~Xu, H.~Liu, L.~Pan, H.~Fan, Z.~Zhang, Z.~Chen, Optimal scheduling of a wind energy dominated distribution network via a deep reinforcement learning approach, Renewable Energy 201 (2022) 792--801.

\bibitem{yang2020deep}
J.~Yang, M.~Yang, M.~Wang, P.~Du, Y.~Yu, A deep reinforcement learning method for managing wind farm uncertainties through energy storage system control and external reserve purchasing, International Journal of Electrical Power \& Energy Systems 119 (2020) 105928.

\bibitem{bucsoniu2010multi}
L.~Bu{\c{s}}oniu, R.~Babu{\v{s}}ka, B.~De~Schutter, Multi-agent reinforcement learning: An overview, Innovations in multi-agent systems and applications-1 (2010) 183--221.

\bibitem{feng2022multi}
L.~Feng, Y.~Xie, B.~Liu, S.~Wang, Multi-level credit assignment for cooperative multi-agent reinforcement learning, Applied Sciences 12~(14) (2022) 6938.

\bibitem{dada2021application}
T.~I. Dada, P.~Thodoroff, N.~D. Lawrence, Application of multi-agent reinforcement learning for battery management in renewable mini-grids, in: AAAI-22 Workshop on Machine Learning for Operations Research (ML4OR), 2021.

\bibitem{karniadakis2021physics}
G.~E. Karniadakis, I.~G. Kevrekidis, L.~Lu, P.~Perdikaris, S.~Wang, L.~Yang, Physics-informed machine learning, Nature Reviews Physics 3~(6) (2021) 422--440.

\bibitem{huang2022applications}
B.~Huang, J.~Wang, Applications of physics-informed neural networks in power systems-a review, IEEE Transactions on Power Systems 38~(1) (2022) 572--588.

\bibitem{stiasny2021learning}
J.~Stiasny, S.~Chevalier, S.~Chatzivasileiadis, Learning without data: Physics-informed neural networks for fast time-domain simulation, in: 2021 IEEE International Conference on Communications, Control, and Computing Technologies for Smart Grids (SmartGridComm), IEEE, 2021, pp. 438--443.

\bibitem{chakraborty2021transfer}
S.~Chakraborty, Transfer learning based multi-fidelity physics informed deep neural network, Journal of Computational Physics 426 (2021) 109942.

\bibitem{banerjee2023survey}
C.~Banerjee, K.~Nguyen, C.~Fookes, M.~Raissi, A survey on physics informed reinforcement learning: Review and open problems, arXiv preprint arXiv:2309.01909 (2023).

\bibitem{gao2022transient}
J.~Gao, S.~Chen, X.~Li, J.~Zhang, Transient voltage control based on physics-informed reinforcement learning, IEEE Journal of Radio Frequency Identification 6 (2022) 905--910.

\bibitem{mikkelsen2003actuator}
R.~Mikkelsen, et~al., Actuator disc methods applied to wind turbines, Ph.D. thesis, PhD thesis, Technical University of Denmark (2003).

\bibitem{annoni2016analysis}
J.~Annoni, P.~M. Gebraad, A.~K. Scholbrock, P.~A. Fleming, J.-W.~v. Wingerden, Analysis of axial-induction-based wind plant control using an engineering and a high-order wind plant model, Wind Energy 19~(6) (2016) 1135--1150.

\bibitem{katic1986simple}
I.~Katic, J.~H{\o}jstrup, N.~O. Jensen, A simple model for cluster efficiency, in: European wind energy association conference and exhibition, Vol.~1, A. Raguzzi Rome, Italy, 1986, pp. 407--410.

\bibitem{li2019constrained}
H.~Li, Z.~Wan, H.~He, Constrained ev charging scheduling based on safe deep reinforcement learning, IEEE Transactions on Smart Grid 11~(3) (2019) 2427--2439.

\bibitem{zhao2022mobile}
H.~Zhao, Z.~Liu, X.~Mai, J.~Zhao, J.~Qiu, G.~Liu, Z.~Y. Dong, A.~M. Ghias, Mobile battery energy storage system control with knowledge-assisted deep reinforcement learning, Energy Conversion and Economics 3~(6) (2022) 381--391.

\bibitem{bordin2017linear}
C.~Bordin, H.~O. Anuta, A.~Crossland, I.~L. Gutierrez, C.~J. Dent, D.~Vigo, A linear programming approach for battery degradation analysis and optimization in offgrid power systems with solar energy integration, Renewable Energy 101 (2017) 417--430.

\bibitem{cortina2017investigation}
G.~Cortina, V.~Sharma, M.~Calaf, Investigation of the incoming wind vector for improved wind turbine yaw-adjustment under different atmospheric and wind farm conditions, Renewable Energy 101 (2017) 376--386.

\bibitem{van2016deep}
H.~Van~Hasselt, A.~Guez, D.~Silver, Deep reinforcement learning with double q-learning, in: Proceedings of the AAAI conference on artificial intelligence, Vol.~30, 2016.

\bibitem{raissi2019physics}
M.~Raissi, P.~Perdikaris, G.~E. Karniadakis, Physics-informed neural networks: A deep learning framework for solving forward and inverse problems involving nonlinear partial differential equations, Journal of Computational physics 378 (2019) 686--707.

\bibitem{boersma2016control}
S.~Boersma, P.~Gebraad, M.~Vali, B.~Doekemeijer, J.~Van~Wingerden, A control-oriented dynamic wind farm flow model:“wfsim”, in: Journal of Physics: Conference Series, Vol. 753, IOP Publishing, 2016, p. 032005.

\bibitem{winddata}
\href{https://www.aedb.org/wind-mast-data}{Wind mast data}, alternative Energy Development Board.
\newline\urlprefix\url{https://www.aedb.org/wind-mast-data}

\bibitem{ma2011grid}
J.~Ma, On-grid electricity tariffs in china: Development, reform and prospects, Energy policy 39~(5) (2011) 2633--2645.

\bibitem{apcchina}
Technical specifications for active power regulation and control of wind farm, Tech. rep., National Energy Administration (2017).

\end{thebibliography}





\end{document}